\documentstyle[12pt,equation,epsfig]{article}
\begin{document}
\newcommand{\be}{\begin{equation}}
\newcommand{\ben}{\begin{subequations}}
\newcommand{\een}{\end{subequations}}
\newcommand{\beq}{\begin{eqalignno}}
\newcommand{\eeq}{\end{eqalignno}}
\newcommand{\ee}{\end{equation}}
\newcommand{\mchi}{\mbox{$m_\chi$}}
\newcommand{\Ochi}{\mbox{$\Omega_\chi h^2$}}
\newcommand{\tanb}{\mbox{$\tan \! \beta$}}
\renewcommand{\thefootnote}{\fnsymbol{footnote}}

\begin{flushright}
APCTP 96--04 \\
September 1996\\
\end{flushright}

\vspace*{2.5cm}
\begin{center}
{\Large \bf Particle Dark Matter}\footnote{To appear in the Proceedings
of the {\it Fourth Workshop on High Energy Physics Phenomenology (WHEPP4)},
Calcutta, India, January 1996.} \\
\vspace{10mm}
Manuel Drees \\
\vspace{5mm}
{\it APCTP, College of Natural Sciences, Seoul National University, 
Seoul 151--742, Korea}
\end{center}
\vspace{10mm}

\begin{abstract}
There is plenty of evidence that most matter in the universe is dark 
(non--luminous). Particle physics offers several possible explanations. In
this talk I focus on cold dark matter; the most promising candidates are then
axions and the lightest supersymmetric particle. I briefly summarize estimates 
for the present relic density of these particles, and describe efforts to
detect them.

\end{abstract}
\clearpage
\setcounter{page}{1}
\pagestyle{plain}
\section*{1) Introduction}
``Dark Matter'' (DM) is matter that does not emit detectable quantities of
electromagnetic radiation; it nevertheless manifests itself through its
gravitational pull. Historical examples are the planets Uranus and Pluto
prior to their discovery by optical telescopes. In this talk I am only
concerned with DM that covers volumes of galactic size or larger. Evidence
for this kind of DM was first collected in the 1920's and 30's, but its
existence became widely accepted only in the 1970's \cite{dr1}. The least
controversial evidence comes from ``galactic rotation curves''. Here one
measures the rotational velocity of globular clusters, hydrogen clouds, or
other objects, around spiral galaxies. Assuming that these objects are
in stable orbits around their parent galaxies, Kepler's law tells us that
the rotational velocity $v_{\rm rot}$ should decrease $\propto 1/\sqrt{r}$
at large distance $r$ to the center of the galaxy, {\em if} the mass of
the galaxy is concetrated in its visible part. However, observationally
all rotation curves become essentially independent of $r$ at large $r$, out
to the largest observable distances; this implies that the mass inside
the radius $r$ grows linearly with $r$, i.e. the mass density drops $\propto
1/r^2$. Cosmologists like to express the average mass density of the 
universe in units of the critical or closure density $\rho_c \equiv
(3 H^2) / (8 \pi G_N)$, where $H$ is the Hubble constant and $G_N$ is
Newton's constant. Numerically, $\rho_c \simeq 2 \cdot 10^{-29} h^2$ g/cm$^3$
$\simeq 1.1 \cdot 10^{-5} h^2$ GeV/cm$^3$, where $h \equiv H/(100$
km/sec$\cdot$Mpc). In these units, galactic DM halos contribute at least
$\Omega \equiv \rho/\rho_c = 0.05$ to 0.1. Note that this is a lower bound, 
since we do not know where these halos end. Indeed, there is considerable
evidence  for larger values of $\Omega$, extending to $\Omega = 1$, from
studies of clusters of galaxies, the ``streaming'' of large numbers of
galaxies, etc \cite{dr2}.

Luminous matter (stars, gas, dust) only contributes $\Omega = 0.01$ or
slightly less. Further, Big Bang nucleosynthesis can only explain the
observed abundances of light isotopes (D, $^3$He, $^4$He, $^7$Li) if
$0.01 \leq \Omega_b h^2 \leq 0.015$, where $\Omega_b$ is the {\em total}
baryonic mass density in the universe. Direct measurements of the Hubble
constant give $0.4 \leq h < 1$. Hence most DM must be non--baryonic,
especially if the total $\Omega = 1$, as favoured by naturalness arguments
and predicted by most inflationary models \cite{dr3}. Note, however, that
this range for $\Omega_b$ also implies the existence of baryonic DM, unless
$h$ is very close to 1. In some sense the recent observation of MACHOs
\cite{dr4} therefore strengthens the argument for non--baryonic DM: MACHOs are
a good candidate for the predicted baryonic matter; having made a
successful prediction obviously makes the overall model that much more
trustworthy.

In order to be able to estimate detection rates of particle DM, we
need to know the {\em local} DM flux, i.e. the local density and
velocity of DM objects. Their density can be estimated by feeding
various observations about our galaxy, including the observed rate of
microlensing (MACHO) events, into a galactic model. A recent study by
Turner et al. \cite{dr5} quotes 0.2 GeV/cm$^3 \leq \rho_{\rm DM}^{\rm
local} \leq $0.5 GeV/cm$^3$ (for a flattened halo, which is the
preferred solution) for the non--MACHO DM density, although an
all--MACHO halo cannot be excluded completely. Note that this is at
least four orders of magnitude larger than the universal DM density
(for $\Omega = 1$).

While the local DM density can at least be constrained from direct 
observations, estimates of the velocity distribution of DM particles are
based almost entirely on galactic modelling. The simplest ansatz for the
DM halo is an isothermal sphere. This leads to a Maxwellian velocity
distribution in the glactic rest frame, with velocity dispersion
$\langle v^2_{\rm DM} \rangle^{0.5} \simeq 270$ km/sec $\simeq 10^{-3} c$.
Note that the solar system moves through this isotropic DM soup with a
velocity of about 220 km/sec. The DM velocity distribution on Earth is
therefore highly anisotropic \cite{dr6}: Most DM particles should come from
the direction in which the solar system is moving. In addition, the Earth
moves around the Sun at about 15 km/sec; this leads to a small annual
modulation of the DM velocity distribution as seen on Earth. Both the
overall anisotropy and the annual modulation can in principle be used to
suppress backgrounds in certain  direct DM searches 
(see sec. 3).\footnote{Recently Cowsik et al. \cite{dr7} suggested that the
velocity dispersion is at least 600 km/sec. However, their analysis has been
criticized by Gates et al. \cite{dr8}. I will therefore stick to the
``canonical'' value for the time being.} Together with the value for
$\rho_{\rm DM}^{\rm local}$ given in the previous paragraph, this gives
a local DM flux $\Phi_{\rm DM} \simeq 10^5 /({\rm cm^2 sec}) \cdot
100 \ {\rm GeV}/m_{\rm DM}$, where $m_{\rm DM}$ is the mass of the DM
particle. For comparison, the flux of cosmic ray muons at sea level is 
about 1/(cm$^2$sec); however, DM particles are much more difficult to
detect than muons.

I focus here on ``cold'' DM (CDM), which was already non--relativistic when
structure formation in the universe began. Scenarios with exclusively
or dominantly ``hot'' (relativistic) DM are essentially excluded
\cite{dr9}, unless the ``seed'' for structure formation comes from exotic
objects like cosmic strings, rather than from quantum fluctuations during
inflation, as commonly assumed.\footnote{There is some evidence that some 
20\% of all DM might be hot, e.g. neutrinos with masses in the eV range
\cite{dr9}. However, other models with only CDM work just as well 
\cite{dr10}.} 
 
Particle physics offers two main classes of CDM candidates,
axions (sec. 2) and weakly interacting massive particles or WIMPs (sec. 3).

\section*{2) Axions}

Let me start by briefly describing axion dark matter. Axions are the
Goldstone bosons of the hypothetical global ``Peccei--Quinn'' (PQ)
symmetry, which has originally been invented to solve the strong CP
problem \cite{dr11}. This new $U(1)$ symmetry allows to rotate the
CP--violating phase $\theta$ away. The PQ symmetry is broken spontaneoulsy
at scale $f_a$, and
explicitly by QCD condensates that break chiral symmetry. This latter
contribution gives the axion a finite mass \cite{dr12}: 
\be \label{e1}
m_a \simeq 6.3 \ {\rm meV} \cdot \frac {10^9 \ {\rm GeV}} {f_a}.
\ee
The couplings of the axion to ordinary matter also scale like the
inverse of $f_a$. Negative laboratory searches can be
translated into upper bounds on these couplings, i.e. lower bounds
on $f_a$. These laboratory bounds require the axion couplings to be
so small that an axion, once produced, can transverse an entire star
without interacting. Axion emission can therefore lead to too rapid
cooling of stars unless the axion production rate is sufficiently small.
The best bounds come from red giants and supernovae; a conservative
limit is \cite{dr12}
\be \label{e2}
f_a > 10^9 \ {\rm GeV} \ \ ==> \ \ m_a \leq 6 \ {\rm meV}.
\ee

An upper bound on $f_a$ can be derived from the requirement that axions do
not overclose the universe. The derivation of this bound is not entirely
straightforward since axions, unlike the WIMPs discussed in the next
section, are not thermal relics. Rather, the main source of axions are
either coherent field oscillations, or axion emission from global (axionic)
strings that are produced during the PQ phase transition. The first
of these sources leads to the bound \cite{dr13}
\be \label{e3}
f_a \leq h^2 \cdot 10^{12 \pm 0.5} \ {\rm GeV} / \langle \theta_a^2
\rangle,
\ee
where $h$ is again the scaled Hubble constant, the uncertainty in the
exponent is due to our limited understanding of details of the QCD phase
transition, and the squared ``vacuum misalignment angle'' 
$\langle \theta_a^2 \rangle$ describes how far the axion field was from
the origin of the potential at the onset of the QCD phase transition.
Notice that the bound (\ref{e3}) becomes very weak if for some
reason $\langle \theta_a^2 \rangle \ll 1$.

The PQ phase transition produces a network of global (axionic)
strings. They are created in a rather energetic state, and
lose energy by emitting axions. Short string loops are found to
be the most potent source of axions. The upper bound on $f_a$ that
results from this source of axions is depicted in Fig. 1, taken
from ref.\cite{dr13}. Here the bound is plotted as a function of the
ratio $\alpha/\kappa$, where $\alpha$ describes the typical length of
string loops, and the ``backreaction parameter'' $\kappa$ describes
the radiation power from strings per unit length; one expects
$0.1 \leq \alpha/\kappa \leq 1$. The bound of Fig. 1 is clearly
stronger than that of eq.(\ref{e3}); however, it can be evaded 
completely in inflationary models, if the reheating temperature is
less than $f_a$. In this case no new strings are produced after
inflation; if any strings were produced before inflation, their
density would be so diluted that they contribute negligibly to total
axion production. In such models the bound (\ref{e3}) would therefore
be the relevant one.

\begin{figure} 
\vspace*{1.5cm}
\centerline{\epsfig{file=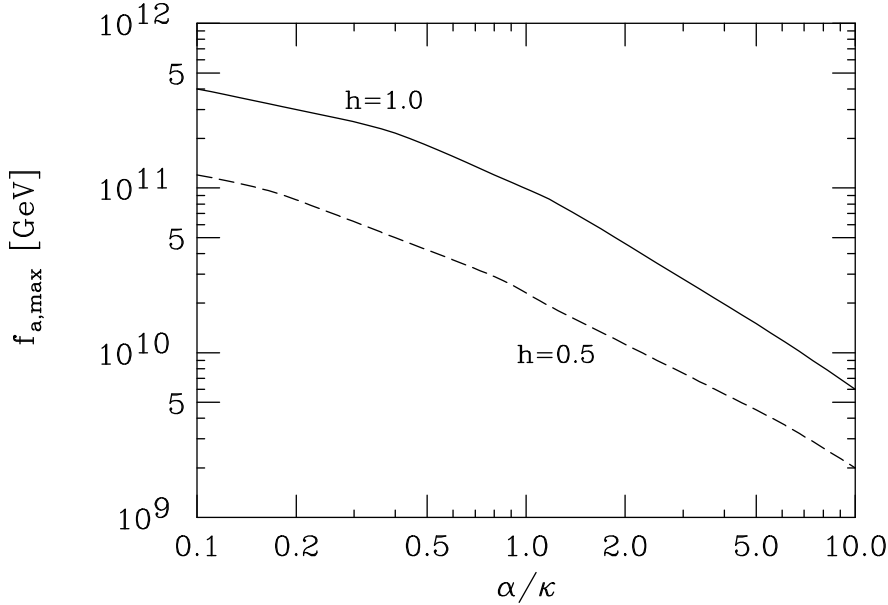,height=6cm}}
\caption 
{The upper bound on $f_a$ is plotted as
a function of the ratio $\alpha/\kappa$ explained in the text, for two
values of the rescaled Hubble constant. It has been assumed that the
present axion density is dominated by axion radiation off cosmic
strings. Adapted from ref.[13].}
\end{figure}

However, the derivation of this bound assumes standard evolution of
the Universe between the QCD phase transition and the present time.
It could be weakened considerably if some mechanism produced a lot
of additional entropy in that period. The reason is that one actually
computes the axion to photon ratio just after the QCD phase transition,
and then scales the photon density to the present one. Clearly this
ratio would be diluted, i.e. the bound on $f_a$ would be weakened, if
an additional source produces a large number of photons after the
QCD phase transition, thereby also increasing the entropy of the
Universe. This is possible in supersymmetric axion models, where the
late decay of the axion's fermionic partner, the ``saxino'', can 
weaken the bound on $f_a$ by as much as a factor $10^3$ \cite{dr14}.
Depending on assumptions, the upper bound on $f_a$ is therefore
somewhere in between a few times $10^{10}$ and $\sim 10^{15}$ GeV;
equivalently, $m_a$ has to be in the range
\be \label{e4}
6 \ {\rm meV} \geq m_a \geq (10^{-2} \ {\rm to} \ 10^2) \ \mu{\rm eV},
\ee
with the narrower window (stronger lower bound) corresponding to more
standard scenarios.

Particles of such a small mass and non--relativistic velocity (see.
Sec.1) cannot be detected by scattering. Rather, one searches for
axion $\rightarrow$ photon conversion in a strong magnetic field
\cite{dr15}. This conversion can occur because 1--loop corrections
produce an $a \gamma \gamma$ coupling of the form
$g_{a \gamma \gamma} a F^{\mu\nu} \widetilde{F}_{\mu\nu}$. The
strength of this coupling is quite model dependent, since even
superheavy particles (with mass $\sim f_a$)
can contribute in the loop. Typically,
\be \label{e5}
g_{a\gamma\gamma} = g_\gamma \frac {\alpha_{\rm em}}
{\pi f_a}, \ \ \ \ \ {\rm with} \ \frac{1}{3} \leq |g_\gamma|
\leq 1.
\ee
This technique is now being pursued \cite{dr16} by an experiment at
Lawrence Livermore National Lab, for axion masses in the $\mu$eV
range. Axions of this mass will convert into microwave photons,
which can be detected only at very low temperatures. Further, a
detectable level of microwave power will only be produced if the
conversion is resonant, i.e. occurs in a cavity whose resonance
frequency satisfies $\omega_{\rm res} = m_a c^2 / \hbar$. In order
to scan a range of axion masses one therefore has to be able to
change $\omega_{\rm res}$ continuously, e.g. by using tuning rods.
The strength of the axion signal (microwave power $P_a$) is then
proportional to \cite{dr16}
\be \label{e6}
P_a \propto g^2_\gamma V B_0^2 \frac{\rho_a}{m_a} Q,
\ee
where $V$ is the volume of the cavity, $B_0$ is the strength of the
applied $B-$field, $\rho_a/m_a$ is the ambient axion number density, and
$Q$ characterizes the quality of the cavity (it is inversely 
proportional to the width of the cavity's resonance). The experiment
now being run at LLNL \cite{dr16} uses $B_0 = 8.5$ T, Q = (a few) $10^5$,
and $V \simeq 0.25 \ {\rm m}^3$. It is expected to probe $|g_\gamma|
\geq 1$ for 1.3 $\mu$eV $\leq m_a \leq$ 13 $\mu$eV. Although
several orders of magnitude more sensitive than earlier pilot
experiments, its sensitivity in terms of both $g_\gamma$ and $m_a$ is
still marginal, see eqs.(\ref{e4}) and (\ref{e5}). In particular, the
covered range of $m_a$ is already excluded in ``standard''
axion cosmology \cite{dr13}, which corresponds to the more stringent
lower bound in (\ref{e3}).

Finally, without going into detail I mention that axions occur quite
naturally in superstring theory \cite{dr17}.

\section*{3) WIMPs}
Unlike axions, weakly interacting massive particles (WIMPs) are
thermal relics. Shortly after the Big Bang they were in thermal
equilibrium with the soup of SM particles, i.e. the WIMP density
was essentially given by the Maxwell--Boltzmann distribution. However,
the rate for reactions that convert WIMPs into SM particles or vice
versa drops quickly once the temperature of the Universe is less
than the WIMP mass \mchi. Eventually this reaction rate will become
smaller than the expansion rate of the Universe, at which
point the WIMPs ``freeze out'', i.e. their density per co--moving
volume remains essentially constant. 

In order to compute the relic WIMP density \Ochi\ one first introduces
the rescaled inverse freeze--out temperature $x_f \equiv \mchi / T_f$;
it satisfies \cite{dr3}
\be \label{e7}
x_f = \ln \frac {0.038 g_{\rm eff} M_p m_\chi 
\langle \sigma_{\rm eff} v \rangle
(x_f)} {\sqrt{g_* x_f}}.
\ee
Here, $g_{\rm eff}$ is the effective number of WIMP degrees of freedom
(e.g., 2 for a single Majorana fermion), $M_P=1.22 \cdot 10^{19}$ GeV is the
Planck mass, and $g_*$ is the effective number of relativistic degrees of
freedom at temperature $T_f$; typically $g_* \simeq 80$ or so. Finally, the
thermal average over the product of the effective WIMP annihilation
cross section $\sigma_{\rm eff} = \sigma(\chi\chi \rightarrow$ anything)
and the relative velocity $v$ between the two annihilating WIMPs is
given by \cite{dr3}
\be \label{e8}
\langle \sigma_{\rm eff} v \rangle
(x) = \frac {x^{1.5}} {2 \sqrt{\pi}} \int_0^\infty dv v^2
e^{-v^2 x/4} \sigma_{\rm eff}(v) \cdot v,
\ee
where I have made use of the fact that WIMPs are non--relativistic at
freeze--out ($x_f \simeq 20$). Eqs.(\ref{e7}) and (\ref{e8}) are
usually solved by numerical iteration. Once $x_f$ has been determined,
the relic density is given by
\be \label{e9}
\Ochi = \frac {1.07 \cdot 10^9 \ {\rm GeV}^{-1}} {J(x_f) 
\sqrt{g_*} M_P},
\ee
where the ``annihilation integral'' $J$ is given by
\be \label{e10}
J(x_f) = \int_{x_f}^\infty \frac {\langle \sigma_{\rm eff} v \rangle (x)}
{x^2} dx.
\ee

In many cases it is sufficient to use a non--relativistic expansion of
the annihilation cross section:
\be \label{e11}
\sigma_{\rm eff} \cdot v = a + b v^2 + \cdots .
\ee
The integrations in eqs.(\ref{e8}) and (\ref{e10}) can then be performed
analytically, with the result \cite{dr3}
\ben \label{e12} \beq
\langle \sigma_{\rm eff} v \rangle (x_f) &= a + \frac {6b}{x_f}
+ \cdots ; \label{e12a} \\
J(x_f) &= \frac {a} {x_f} + \frac {3b} {x_f^2} + \cdots . \label{e12b}
\eeq \een
However, this approximation breaks down if $\sigma_{\rm eff}$
depends sensitively on $v$, e.g. in the vicinity of $s-$channel
poles or near a threshold. Further, in some cases the WIMP stays in
{\em relative} thermal equilibrium with another, slightly heavier new
particle $\chi'$ even after it has dropped out of equilibrium with SM
particles. In this case ``co--annihilation'' reactions of the form
$\chi \chi' \rightarrow$ (SM particles) have to be included into the
calculation of $\sigma_{\rm eff}$.
I refer the reader to ref.\cite{dr18} for a discussion of these more
complicated cases.

Eq.(\ref{e7}) shows that the freeze--out temperature depends only
logarithmically on the annihilation cross section. All potentially realistic
WIMP candidates therefore have $x_f \simeq 20$, which corresponds to
$v \simeq 1/3$, or $J(x_f) \simeq \sigma_{\rm eff}v/20$ if the expansion
(\ref{e11}) can be used. Plugging this into eq.(\ref{e9}) one finds
that \Ochi\ is of order unity, i.e. $\chi$ makes a good CDM candidate, if
\be \label{e13}
\sigma_{\rm eff} \simeq 2 \cdot 10^{-10} \ {\rm GeV}^{-2}
\sim {\cal O}(0.1) \ {\rm pb},
\ee
which is similar to typical weak interaction cross sections! This is a
highly nontrivial ``coincidence''; notice, e.g., the factors of
$M_P$ in eqs.(\ref{e7}) and (\ref{e9}), which somehow conspire with
other numerical factors to single out the weak scale. It means that
we do not have to invent either very strong or very weak new interactions
in order to explain the existence of Dark Matter; any stable particle
with ``ordinary'' weak interactions will do. (Hence the WI in WIMP, of 
course.)

A WIMP that can annihilate into SM particles with roughly weak interaction
strength usually also scatters off ordinary matter with roughly weak
interaction cross sections. The two most promising WIMP search techniques
make use of these interactions.

``Direct'' WIMP searches look for energy deposited by a WIMP 
inside a detector in a laboratory (situated underground to reduce
cosmic ray induced backgrounds). Given that ambient CDM particles are
expected to have velocity $v_\chi \simeq 10^{-3} c$, see Sec.~1, the
deposited energy can only be of order $E_{\rm vis} \leq 50 \ {\rm keV}
\cdot \mchi / (100 \ {\rm GeV})$. Such a small energy deposition can be
detected calorimetrically only if the detector is cooled.
This technique was employed by the first experiments, using Germanium
detectors that were originally designed to search for neutrino--less
double $\beta$ decay \cite{dr19}. However, it is difficult to use
this technique for detectors weighing more than a pound or so. More
recently, people have therefore started to employ scintillating
detectors, e.g. NaI. The currently most stringent direct detection
limits come from such devices \cite{dr20}; they restrict the WIMP
scattering rate to be below a few tens of events/(kg$\cdot$day).
Future improvements are expected when the detectors are cooled to
liquid nitrogen temperature. This should improve the discrimination
between backgrounds due to $\beta$ and $\gamma$ radioactiv decays in
or near the detector, and the signal due to nuclear recoil, which
produces a slightly different light curve in the scintillator
\cite{dr21}. Other scintillating materials are also being explored;
liquid Xenon appears to be particularly promising for large scale
detectors \cite{dr22}.

Direct WIMP detection becomes difficult if \mchi\ greatly exceeds
100 GeV. On the one hand, the WIMP flux decreases like $1/\mchi$
for given ambient CDM mass density. Further, for heavy WIMPs and
not very small nuclei the momentum transfer can be large enough to
lead to significant nuclear form factor suppression of the scattering
cross section. Fortunately the second promising WIMP search strategy
actually works better for heavier WIMPs.

This ``indirect detection'' technique is based on the observation
\cite{dr23} that WIMPs can be trapped inside the Sun or Earth, if
they lose a sufficient amount of energy in a scattering reaction
while travelling through these celestial bodies to become gravitationally
bound to them. Such WIMPs will become concentrated near the center
of these bodies. Eventually equilibrium will be reached between the
rate of WIMP annihilation in and WIMP capture by these bodies, i.e.
the annihilation rate becomes one half the capture rate (note that
each annihilation destroys two WIMPs). Most annihilation
products will be absorbed immediately in the surrounding medium.
However, neutrinos can escape, and might then be detected in 
underground neutrino detectors on Earth. The by far largest signal
comes from muon--neutrinos, since they can be converted into a muon
well outside the detector proper, thereby greatly increasing the
effective target volume. Furthermore, muons point back into the
direction of the parent neutrino. One can therefore enhance the
sensitivity of the search by comparing the neutrino flux from the center
of the Sun or Earth with side bins. The expression for this indirect
WIMP detection rate also receives a factor $1/\mchi$ from the CDM flux.
However, for $\mchi < 1$ TeV this suppression is more than compensated
by two enhancement factors: The $\nu_\mu \rightarrow \mu$ conversion
cross section rises essentially linearly with energy (as long as 
$m_p \mchi \ll m_W^2$, where $m_p$ is the proton mass); and the
range of muons produced in the rock (or ice) surrounding the
detector, and hence the effective detector volume for $\nu_\mu$'s
coming from a fixed direction, increases linearly with $E_\mu \propto
\mchi$. Increasing \mchi\ beyond a TeV or so decreases the rate
again, mostly because it becomes less likely for very heavy
WIMPs to become trapped gravitationally after scattering off a
nucleus in the Sun or Earth.

Both the direct and the indirect WIMP detection rate scale linearly
with the WIMP--matter scattering cross section.\footnote{In case of
indirect detection this is true only if equilibrium between capture
and annihilation has been reached. Prior to equilibrium the signal
is exponentially suppressed and therefore essentially unobservable
\cite{dr23}; this is the case for the signal from the Earth's center
for most WIMP candidates that are significantly heavier than iron
nuclei.} Since ambient CDM particles are non--relativistic, this
cross section grows with the available center--of--mass energy
$\sqrt{s}$. Scattering off electrons can therefore almost always be
neglected compared to scattering off nuclei. One then distinguishes
between spin--dependent interactions, through a pseudoscalar or
axial vector coupling to the spin of the nucleus in question, and
spin--indenpedent scalar or vector interactions. The latter can
couple coherently to an entire nucleus, leading to an extra
enhancement factor $A^2$ compared to the former, where $A$ is the
nucleon number of the target nucleus. As already noted above, if
both the WIMP and the nucleus are heavy, nuclear form factor
suppression has usually to be taken into account \cite{dr6}. The best bounds on
WIMPs with dominantly spin--dependent interactions comes from the
NaI experiment mentioned above \cite{dr20}. The best current bounds on
WIMPs with mostly spin--independent couplings come from indirect
searches at the underground detectors Kamiokande \cite{dr24} and
Baksan \cite{dr25}.

The perhaps most obvious WIMP candidate is a heavy neutrino. However,
this suffers from both theoretical and experimental problems. 
Theoretically, it is not at all clear why a heavy neutrino should be
stable, given that all known heavy SM fermions decay very rapidly.
Further, an $SU(2)$ doublet (Dirac or Majorana) neutrino would have
been found by existing searches, if its mass exceeds about 10 GeV
(which is required by LEP data), and if it constitutes a substantial
fraction of the DM halo of our galaxy. Actually, such a doublet neutrino
does {\em not} make a good DM candidate unless it is very heavy; its
relic density is too low, due to strong annihilation into $W$ and
$Z$ pairs, unless its mass is in the TeV range. One way out is to
introduce an $SU(2)$ singlet fermion, which mixes with the doublet 
after $SU(2)$ is broken. If the lighter eigenstate is stable,
one can arrange the mixing angle such that $\Ochi \simeq 1$ as desired.
Such models are not only contrived, they are also severly
constrained by existing data \cite{dr20,dr24,dr25}.

The currently by far most popular WIMP candidate is therefore the
lightest supersymmetric particle (LSP) \cite{dr2}. If one only includes
those terms in the Lagrangian that are necessary to produce quark and
lepton masses, $R-$parity is conserved and the LSP is stable. It then
has to be electrically neutral in order to avoid bounds from exotic
isotope searches \cite{dr27}. This leaves us with two candidates, a 
sneutrino and the lightest neutralino. Sneutrinos have similar annihilation
cross sections and capture rates as Dirac neutrinos; they are therefore
essentially excluded as WIMP candidates \cite{dr28}. 

Neutralino dark matter has been studied quite extensively \cite{dr2}.
In the MSSM the neutralinos are mixtures of the bino, the neutral wino, and 
the two neutral higgsinos; extended models also include singlet
higgsinos \cite{dr29}, but I will stick to the MSSM here. Under the usual
assumption of gaugino mass unification the LSP can never be wino--like,
since in this case the $SU(2)$ gaugino mass $M_2$ is about twice as 
large as the $U(1)$ gaugino mass $M_1$ at the weak scale. One thus has to
consider three scenarios. If the higgsino mass parameter
$|\mu| \gg M_1$, the LSP is mostly a gaugino (a photino if $M_1 \ll m_Z$,
and mostly bino if $M_1 \geq m_Z$). In the opposite limit $M_1 \gg |\mu|$,
the LSP is mostly higgsino. Finally, if $M_1 \simeq |\mu|$, the LSP is
a strongly mixed state.

In order to decide whether the LSP makes a good CDM candidate one first
has to estimate its relic density. The to date most complete list of
LSP annihilation cross sections has been compiled in ref.\cite{dr30}.
Here all possible 2--body final states have been treated, at tree level
and using the expansion of eqs.(\ref{e11}), (\ref{e12}). It was realized
subsequently that this is not always sufficient for a reliable estimate
of the relic density. First, if the LSP is higgsino--like, the lighter
chargino and the next--to--lightest neutralino are also higgsino--like 
and only slightly heavier than the LSP. In this case co--anihilation
between the LSP and these slightly heavier states can be important.
This is true in particular for $|\mu| \leq m_W$. The reason is
that the $Z \widetilde{\chi}_1^0 \widetilde{\chi}_1^0$ coupling is
very small for almost pure higgsinos, while the $Z 
\widetilde{\chi}_2^0 \widetilde{\chi}_1^0$ and $W \widetilde{\chi}_1^\pm
\widetilde{\chi}_1^0$ couplings have essentially full gauge strength.
As a result, light higgsinos become uninteresting as CDM candidates
\cite{dr31}.

Furthermore, $s-$channel exchange diagrams involving Higgs bosons are
often quite important. In this case the non--relativistic expansion
(\ref{e11}) becomes unreliable. In this expansion one assumes that
LSPs annihilate essentially at rest, i.e. the center--of--mass
energy $\sqrt{s} = 2 \mchi$. However, the thermal energy of the LSPs
can often push $\sqrt{s}$ up to the mass $m_H$ of the exchanged Higgs
boson if 2\mchi\ is not too much smaller than $m_H$, thereby greatly
reducing \Ochi. Not surprisingly, this effect is especially important
in models where other ($t-$ and $u-$ channel) contributions to the
annihilation cross section are small \cite{dr32}.

The dashed curves in Figs. 2 and 3 show contours of constant \Ochi\
including some co--annihilation effects, and using a careful treatment
of $s-$channel exchange contributions following ref.\cite{dr33}. For
these plots I have assumed gaugino mass unification, and have used a
common mass $m_0$, as well as a common $A-$parameter $A_0$, for all 
sfermions at the GUT scale $M_X = 2 \cdot 10^{16}$ GeV. At the weak scale
squarks are then heavier than sleptons (by as much as a factor 3--5),
and $SU(2)$ doublet sleptons are heavier than $SU(2)$ singlets. 
However, I have left the SUSY breaking contributions to the Higgs boson
masses free. This allows me to treat the mass $m_A$ of the pseudoscalar
Higgs boson as well as $\mu$ as free parameters. I have taken $\tanb=2,
\ m_t =170$ GeV, $A_0=0$ and $m_A=500$ GeV; in Fig.~2, $m_0=200$ GeV,
while Fig.~3 is for $m_0=400$ GeV. 

In both figures the lower--left region of the $(\mu,M_2)$ plane is
excluded by LEP searches for charginos and neutralinos. The lower--right
corners are excluded since here the lighter stop eigenstate $\tilde{t}_1$
would be lighter than the lightest neutralino; note that the
off--diagonal entries of the stop mass matrix grow $\propto \mu$. The
region excluded by this constraint is larger in Fig.~2 since it has
smaller diagonal stop masses (smaller $m_0$). Finally, in the
top--right corner of Fig.~2, the lighter stau eigenstate 
$\widetilde{\tau}_1$ becomes lighter than the lightest neutralino. Note
that the ratio of stop and neutralino masses increases with increasing
$M_2$ since $m_{\tilde{t}_1}$ gets large positive contributions from
gluino loops. In contrast, the stau to neutralino mass ratio decreases
with increasing $M_2$, since $m_{\widetilde{\tau}_1}$ only receives
small contributions from bino loops.

Figs. 2 and 3 show prominent effects from the $H^0$ and $A$ exchange
poles at $M_2 \simeq 500$ GeV, and from the merged $Z$ and $h^0$ 
exchange poles at $M_2 \simeq 100$ GeV; here $h^0$ and $H^0$ are
the light and heavy neutral Higgs scalars, respectively. For
$m_0=200$ GeV, Fig.~2, only a small region of the plane (top right)
is excluded by the requirement $\Ochi \leq 1$. Increasing $m_0$ to
400 GeV, Fig.~3, enlarges this region dramatically. The reason is that
gaugino--like LSPs annihilate dominantly through $t-$channel exchange
of $SU(2)$ singlet sleptons, which have both the smallest mass and the
largest hypercharge of all sfermions. In the gaugino region the
relic density therefore depends sensitively on $m_0$, roughly
$\propto m_0^4$ if $M_1^2 \ll m_0^2$. On the other hand, in the
higgsino region, $|\mu| < M_1 \simeq M_2/2$, the relic density depends
very little on $m_0$. For $|\mu| < m_W$ co--annihilation into
fermion--antifermion pairs via $W$ and $Z$ exchange is the dominant
process, while heavy higgsinos annihilate dominantly into $W$ or $Z$
pairs; neither of these reactions depends on sfermion masses.

\begin{figure} 
\vspace*{2.cm}
\centerline{\epsfig{file=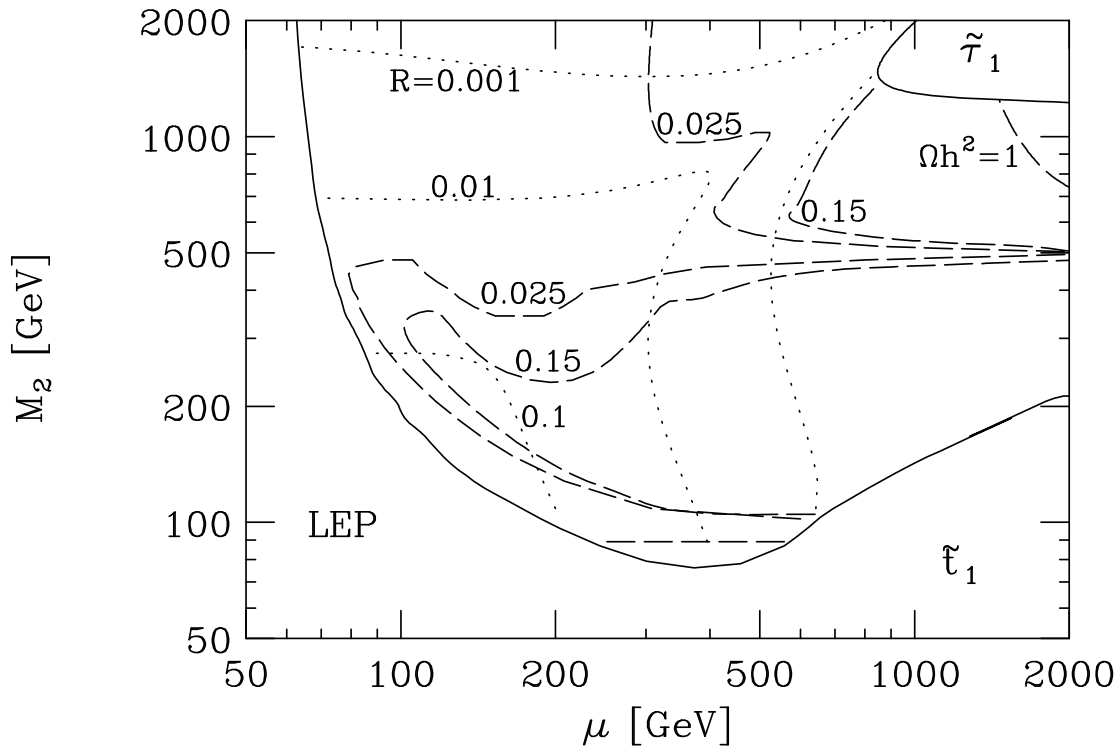,height=7.5cm}}
\caption 
{Contours of constant relic density (dashed), and of constant scattering
rate in $^{76}Ge$ (dotted, in units of evts/(kg$\cdot$day)). The values of
the free parameters are: $m_t=170$ GeV, $\tan \! \beta = 2, \ A_0 = 0,
m_A=500$ GeV and $m_0=200$ GeV. The regions outside the
solid curves are excluded, as described in the text.}
\end{figure}

\begin{figure} 
\vspace*{2.cm}
\centerline{\epsfig{file=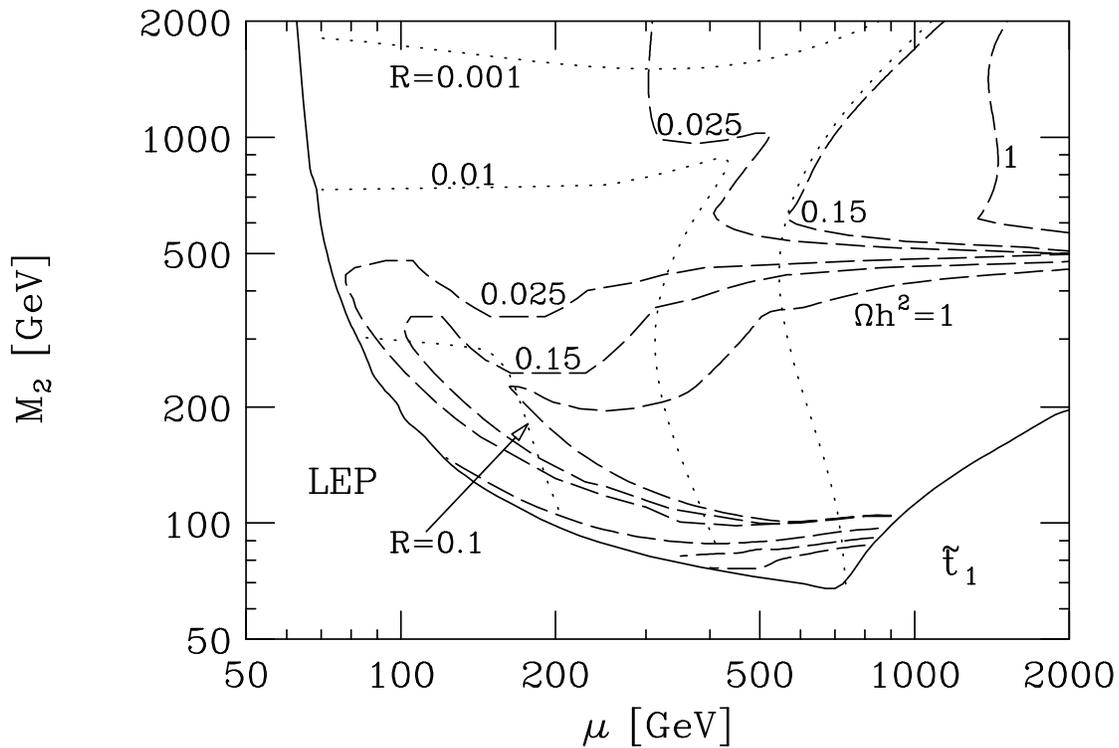,height=7.5cm}}
\caption 
{As Fig.~2, but for $m_0=400$ GeV.} 
\end{figure}

However, the estimate of the relic density in the higgsino region could
be changed significantly by two effects that are not taken into account
in these figures. First, co--annihilation into states other than SM
fermions or a photon and a $W$ boson have not been included. Heavy
higgsinos can co--annihilate into a host of other final states involving
gauge or Higgs bosons; this will reduce their relic density even more.
Secondly, it has recently been pointed out \cite{dr34} that one--loop
corrections can significantly change the mass splitting between
higgsino--like states. These corrections can have either sign.
If they increase the mass splitting, co--annihilation
would be suppressed (its rate depends exponentially on the mass
splitting), possibly reinstating light higgsinos as viable WIMP
candidates.

Fig.~2 shows that a gaugino--like LSP usually does make a good CDM
candidate, provided that $m_0$ is not too far from 200 GeV. In this
context it is interesting to note that minimal supergravity models
with heavy top quark favour a gaugino--like LSP, although a 
higgsino--like LSP remains possible even in these restrictive models.
By ``minimal supergravity'' I mean models where the universality
of soft breaking scalar masses is extended into the Higgs sector.
The electroweak gauge symmetry is then more or less automatically
broken by radiative corrections \cite{dr35}, which allows to determine
$|\mu|$ and $m_A$ in terms of $m_t, \ \tanb, \ M_2$ and $A_0$. One
might hope that in these restrictive models the requirement $\Ochi 
\leq 1$ allows to place upper limits on SUSY breaking parameters.
Unfortunately this is not the case \cite{dr30}, as illustrated in
Fig.~4. The solid curve in this plot shows \Ochi\ as a function of
\tanb\ for $m_0=500$ GeV, $A_0=0, \ m_t=170$ GeV, $M_2=250$ GeV
and $\mu<0$. For small \tanb\ the relic density is indeed
unacceptably high, due to the large value of $m_0$ chosen here.
However, two effects reduce \Ochi\ as \tanb\ is increased. First,
the lighter $\widetilde{\tau}$ eigenstate becomes lighter \cite{dr36}, due to
loop corrections involving the $\tau$ Yukawa coupling (which grows
with \tanb), and also due to increased $\widetilde{\tau}_L - 
\widetilde{\tau}_R$ mixing. Secondly, and even more importantly, 
$m_A$ becomes smaller \cite{dr36}, due to loop effects involving the
$b$ and $\tau$ Yukawa couplings. For $\tanb \simeq 48, \ 2 \mchi \simeq
m_A$ and the relic density is very small. One is thus forced to conclude
that imposing an upper bound on the
LSP relic density does not lead to strict upper bounds on sparticle
masses, although it does significantly constrain the allowed parameter
space.

\begin{figure} 
\vspace*{2.cm}
\centerline{\epsfig{file=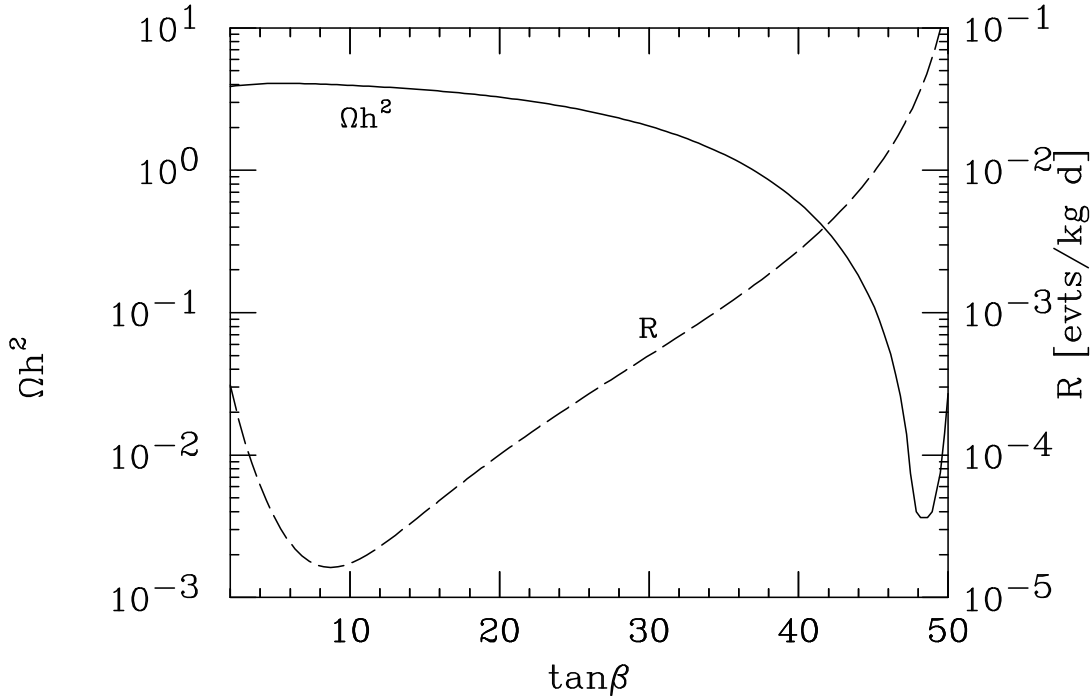,height=7cm}}
\caption 
{The LSP relic density (solid, referring to scale at left) and detection
rate in $^{76}$Ge (dashed, referring to scale at right) are plotted as
a function of $\tan \! \beta$ in a minimal supergravity model witgh radiative
gauge symmetry breaking. The values of the other free parameters are $M_2=250$
GeV, $m_t=170$ GeV, $A_0=0$ and $\mu<0$.} 
\end{figure}

Let me finally briefly discuss LSP detection. As mentioned earlier, both the
direct and the indirect detection rate scale with the LSP--matter
scattering cross section. The to date most comprehensive calculation of
this cross section has been performed in ref.\cite{dr37}. Since
neutralinos are Majorana fermions, their vector current vanishes
identically and their pseudoscalar current is proportional to the LSP
velocity $\sim 10^{-3}$. This means that $Z$ exchange only contributes
to spin--dependent interactions, while pseudoscalar Higgs exchange can
be neglected entirely. The exchange of scalar Higgs bosons gives rise to
spin--independent interactions, and in fact often dominates the total
scattering cross section. Finally, squark exchange contributes to both 
spin--dependent and spin--independent interactions. However, in models
with gaugino mass unification and universal scalar masses at the GUT
scale one has $m_{\widetilde{q}} \geq 5 \mchi$, in which case squark
exchange contributions are quite small.\footnote{The lighter stop
eigenstate can be close to the LSP mass in such models, as noted
above. However, there are no top quarks in the proton, so $\tilde{t}_1$
exchange can only contribute at 1--loop level, and remains quite
small \cite{dr37}.}

As an example, the dotted lines in Figs. 2 and 3 show contours of
constant scattering rate in $^{76}$Ge detectors in units of
events/(kg$\cdot$day), ignoring possible energy detection thresholds
(which would reduce the rate), and assuming that the galactic halo is
formed essentially entirely by LSPs. Recall that the present bounds on
this rate lie above 10 evts/(kg$\cdot$day). Clearly even increasing the
sensitivity by a factor 100 only begins to scratch the depicted parameter
space. Note also that in much of the region giving relatively large
scattering rates for {\em fixed} local DM density the LSP does in fact
not make a good CDM candidate, since \Ochi\ is too low to even produce 
galactic haloes, which requires $\Ochi \geq 0.025$ or so. The
situation is even worse for $\mu < 0$, as shown by the dashed line in Fig.~4:
Although $M_2 =250$ GeV is quite moderate, the detection rate can be
some five orders of magnitude below the present bound, due to destructive
interference between $h^0$ and $H^0$ exchange. At large \tanb\ the
rate increases, because the couplings of $H^0$ to $d$ and $s$ quarks
grow $\propto \tanb$, and also because $m_{H^0} \simeq M_A$
becomes smaller, as mentioned earlier. However, the mass of the charged
Higgs boson is also reduced, which can lead to conflict with the
upper bound on the branching ratio for $b \rightarrow s \gamma$ decays.
Indeed, in ref.\cite{dr38} it was pointed out that this last constraint
makes it very difficult to construct viable SUSY models with large
LSP detection rate even if all parameters are set ``by hand'' at the
weak scale.\footnote{The authors of ref.\cite{dr39} did manage to
do just that. However, they had to chose squark masses just above \mchi\
and $m_A$ well below \mchi; neither assumption is particularly natural
from a model building point of view.}

\section*{4) Summary and Conclusions}
The argument for the existence of some exotic dark matter is quite
compelling. Studies of structure formation in inflationary models
indicate that most of this dark matter should have been ``cold''
already at the epoch of galaxy formation. Particle physics conveniently
provides us with (at least) two CDM candidates: The axion (if $f_a$ is chosen
appropriately), and the
supersymmetric LSP (if gaugino--like). Both have the attractive feature
that they were originally introduced to solve a problem that has nothing
to do with dark matter: Axions solve the strong CP problem, and supersymmetry
solves the hierarchy problem. We have seen that both candidates are quite
difficult to detect (other than through their gravitational effects);
experiments of the present and even next generations will only begin to
probe the allowed parameter space. One should keep in mind that things
could be even worse, though. For example, CDM could come from the ``hidden
sector'' present in many supergravity or superstring models, in which
case it would interact with normal matter {\em only} gravitationally,.
Individual CDM particles would then be completely undetectable.

However, before we have to consider such depressing scenarios, we should
fully explore the parameter space of the more attractive detectable
candidates. In case of the LSP, a first decisive test will probably come
from the LHC (or, if we are lucky, from LEP). However, even if the LHC
finds a SUSY signal, detection of relic LSPs is necessary to
prove that they are stable on cosmological time scales. (Collider
experiments can only establish a lower bound of $10^{-8}$ seconds or so.)
In contrast, I do not know of any realistic scheme to detect axions
produced in the lab, given the extremely strong upper limits on their
interaction strength; CDM axions are our only hope to detect these
elusive particles, if they indeed exist. This once again illustrates the
increasingly tight connection between particle physics and cosmology.

\end{document}